\documentstyle[prl,multicol,aps]{revtex}
\def\sech{\mbox{sech}}

\begin{document}
\input{epsf.tex}
\epsfverbosetrue
\title{Multi-soliton energy transport in anharmonic lattices}

\author{Elena A. Ostrovskaya$^1$, Serge F. Mingaleev$^{1,2,3}$,
Yuri S. Kivshar$^1$, Yuri B. Gaididei$^{2,3}$, and 
Peter L. Christiansen$^2$}

\address{$^1$ Optical Sciences Centre, Australian
National University, Canberra ACT 0200, Australia \\
$^2$ Department of Mathematical Modeling, Technical University of
Denmark, Lyngby DK-2800, Denmark \\
$^3$ Bogolyubov Institute for Theoretical Physics, 
Kiev UA-03143, Ukraine}

\newcommand{\B}{${\mathcal B}$}
\maketitle

\begin{abstract}
We demonstrate the existence of {\em dynamically stable multihump 
solitary waves} in polaron-type models describing interaction 
of envelope and lattice excitations. In comparison with
the earlier theory of multihump optical solitons 
[see Phys. Rev. Lett. {\bf 83}, 296 (1999)], our analysis reveals 
{\em a novel physical mechanism} for the formation of stable 
multihump solitary waves in nonintegrable multi-component 
nonlinear models.
\end{abstract}

\pacs{PACS numbers: 03.40.Kf}

\begin{multicols}{2}
\narrowtext


Spatially localised solutions of multi-component 
nonlinear models, {\em multi-component solitary waves}, 
have received a great deal of attention in the last decade. In 
particular, recent studies in the nonlinear optics 
\cite{moti,ours} and Bose-Einstein condensation \cite{bec} 
have shown that, under certain conditions and only in 
multi-component systems, the formation of dynamically stable 
 localized states and soliton complexes is possible. Unlike 
their single-component (or {\em scalar}) counterparts, multi-component 
(or {\em vector}) solitons possess complex internal structure 
forming a kind of ``soliton molecules'', 
which makes them attractive, both from the fundamental 
and applied point of view, as composite and reconfigurable 
carriers for a transport of spatially localized energy. 

Recent discovery of {\em stable multi-component} spatial 
solitons in optics \cite{moti,ours} shed a light on the general 
physical mechanisms of the formation and {\em stability} of multi-component 
localised states. Such states are often called {\em multihump solitons} due to multiple maxima displayed in their intensity profile. Usually, multihump solitary waves appear via 
{\em bifurcations} of scalar solitons when a primary soliton plays 
a role of an effective waveguide (``potential well'') that 
traps higher-order guided 
modes excited in a complimentary field \cite{ours}. On 
the other hand, the multihump solitons can be formed as 
{\em multi-soliton bound states}, when two or more different 
vector solitons are ``glued'' together due to balanced interaction 
between the soliton constituents \cite{gluons}. 

{\em Soliton bifurcations and binding} enable the existence of
multi-component localized states in many nonlinear models; these include bound 
states of dark solitons \cite{gluons} and incoherent solitons 
\cite{incoh} in optics, and multihump plasma waves \cite{Kol'chugina}. Importantly, multihump solitary waves are also found in 
higher dimensions \cite{Kol'chugina,moti2}. 

The experimental and theoretical results on optical
solitons \cite{moti,ours} challenge the conventional view on
multi-component solitary waves in other fields of nonlinear physics.
The main question we wish to address here is: {\em Can stable multihump
solitons exist in other important models of nonlinear physics}?
This is {\em a crucial issue} because, so far, the stable multihump 
solitons have been positively identified only in the nonlinear 
optical model of Refs. \cite{moti,ours} that is known to possess additional 
symmetries, which might be the reason for their unique stability.

In this Letter, we demonstrate {\em the existence of dynamically 
stable multihump solitary states} in a completely different (in 
both the physics and properties) but {\em even more general model} 
that describes the interaction of 
envelope and lattice excitations, a generalisation of the well-known
{\em polaron model}. We reveal a 
{\em novel physical mechanism} for the formation of stable multihump 
solitary waves in nonintegrable multi-component nonlinear models.

{\em Model.}
Let us consider the continuous model of the energy 
(or excess electron) transport in an anharmonic molecular 
chain \cite{Dav-book}, described by the system of 
coupled nonlinear Schr{\"o}dinger (NLS) and Boussinesq equations:
\begin{eqnarray}\label{pde}
i \frac{\partial \psi}{\partial t} &+& \frac{\partial^2 
\psi}{\partial x^2} + 2 w \psi =0 \; , \nonumber \\ 
m \frac{\partial^2 w}{\partial t^2} &=& 
\frac{\partial^2 w}{\partial x^2} + \mu 
\frac{\partial^4 w}{\partial x^4} + 
\alpha \frac{\partial(w^2)}{\partial x^2} - 
\frac{\partial(|\psi|^2)}{\partial x^2} \; , 
\end{eqnarray}
where $t$ and $x$ are the normalised time and spatial 
coordinate, correspondingly, $\psi(x,t)$ is the 
excitation wave function, and $w(x,t)$ is the chain strain. 
The system is characterized by three
dimensionless parameters: the particle mass $m$, the anharmonicity of
the chain $\alpha$, and the dispersion coefficient $\mu$.

Equation (\ref{pde}) appears in a number of other 
physical contexts including, for example, the interaction of 
nonlinear electron-plasma and ion-acoustic waves 
\cite{Nishikawa}, coupled Langmuir and ion-acoustic plasma waves 
\cite{Makhankov}, interaction  of optical and acoustic modes in 
diatomic lattices \cite{Yajima}, particle 
 theory models \cite{Brown}, etc. 

System (\ref{pde}) is known to be integrable for 
$\alpha \mu = 6$ \cite{Krichever}. 
In this case, it possesses two types of single-soliton 
solutions: scalar ($\psi=0$) supersonic Boussinesq (\B q) solitons 
and vector Davydov-Scott (DS) solitons \cite{Dav-book} which can 
be both {\em  subsonic} and {\em supersonic}. Because 
of the complete integrability for $\alpha \mu = 6$, these solitons do not
interact with each other. For $\alpha \mu \neq 6$, the situation 
changes dramatically, and it has been recently shown 
\cite{Gaididei,Zolotaryuk} 
for the nearly-integrable case ($\alpha \mu \approx 6$) 
that \B q and DS solitons can form a bound state 
for  $\alpha \mu > 6$. From the other hand, 
it is also known that in a weakly
anharmonic lattice two subsonic DS solitons can form a {\em bisoliton} 
\cite{Brizhik}. However, it remains a mystery 
what happens when the system (\ref{pde}) is far from its 
integrable limit and, especially, when the solitons are supersonic.
In this Letter we examine the model (\ref{pde}) numerically 
for arbitrary values of $\alpha \mu$ and, employing the concept of 
soliton bifurcations, demonstrate the origin and exceptional 
robustness of {\em multihump supersonic} stationary solitary waves. 

\begin{figure}
\setlength{\epsfxsize}{5.7 cm}
\centerline{\epsfbox{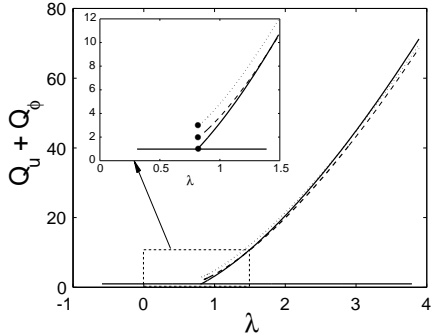}}
\caption{Bifurcation diagram of the system (\ref{ode}) 
for $g=7$. Inset - close-up of 
the bifurcation point. Horizontal line - $u$-solitons, solid -
single-hump $|0,0\rangle$ solitons, dashed - two-hump 
solitons, dotted - three-hump solitons.}
\label{sup_bif}
\end{figure}
\begin{figure}
\setlength{\epsfxsize}{7.0 cm}
\centerline{\epsfbox{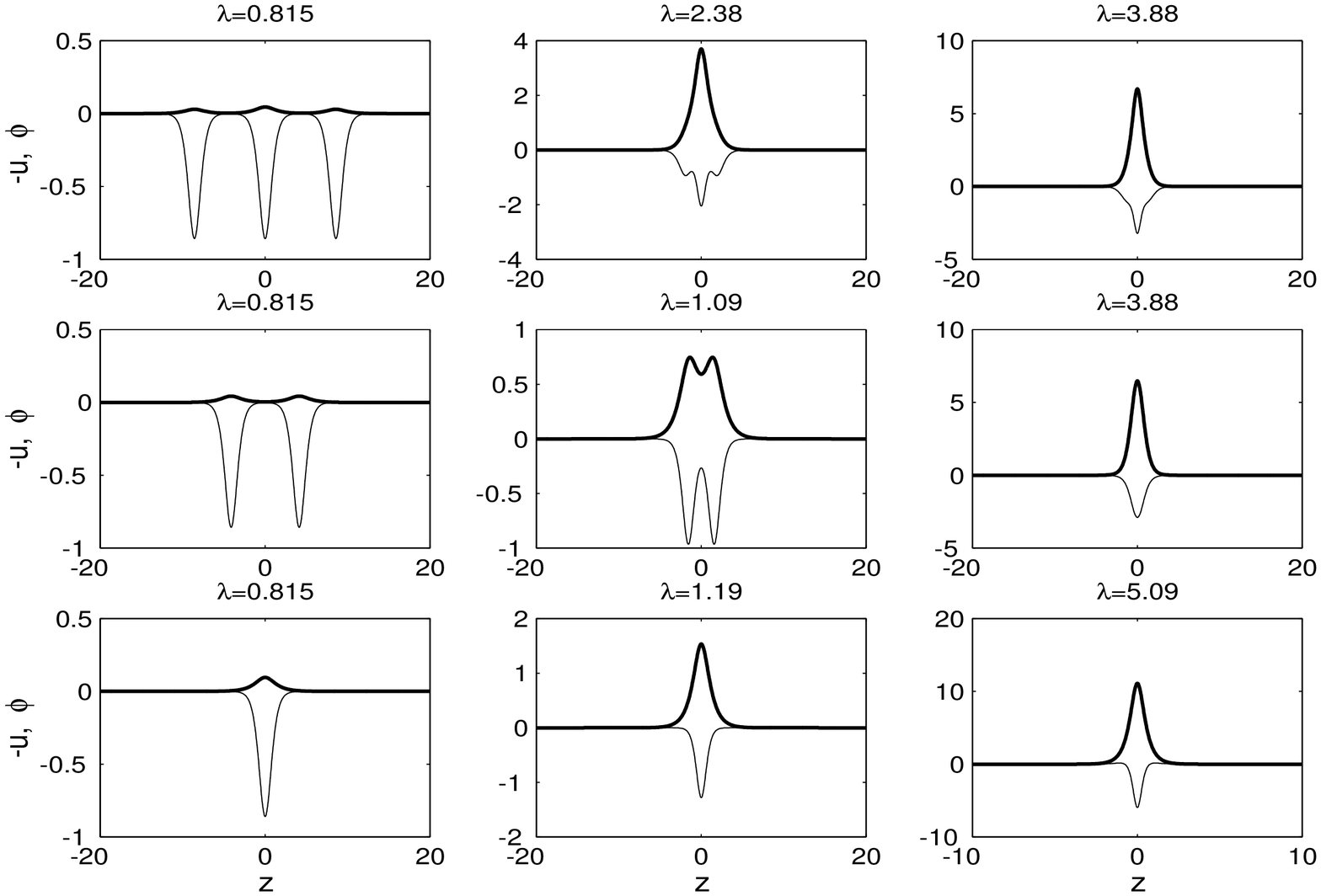}}
\caption{Profiles of single-hump (bottom row), two-hump (middle row),
and three-hump (top row) solitons with increasing (from left to right) 
values of $\lambda$.}
\label{sup_prof}
\end{figure}

{\em Bifurcation analysis.} 
The stationary solutions of Eq. (\ref{pde}) 
can be found in the form of the traveling waves
\begin{eqnarray}\label{scale}
\psi(x,t)= \frac{c}{4\sqrt{\mu}} e^{ivx/2-i\Omega t} \phi(z) \; ,
\quad 
w(x,t)= \frac{c}{4\mu} u(z) \; ,
\end{eqnarray} 
where $z=\sqrt{c/4\mu} (x-vt)$, and the constant $c=(mv^2-1)$ is 
positive for a supersonic velocity, $v>1/\sqrt{m}$. 
Substituting Eq. (\ref{scale}) into Eq. (\ref{pde}), we derive 
a system of coupled ordinary differential equations 
\begin{eqnarray}\label{ode}
\phi_{zz} - \lambda \phi + 2 u \phi = 0 \; , \nonumber \\
u_{zz} -4 u + g u^2 - \phi^2 = 0 \; ,
\end{eqnarray} 
where $g=\alpha \mu$ is an effective anharmonicity parameter, 
and $\lambda = \mu (v^2-4\Omega)/c$ is a characteristic eigenvalue 
of the stationary localized solutions.

Equation (\ref{ode}) has two types of one-soliton 
solutions: {\em a one-component \B q soliton}
\begin{eqnarray}\label{bq}
\phi_0(z)=0 \; , \qquad 
u_0(z)=(6/g) \, \sech^2(z) \; , 
\end{eqnarray} 
which exists for arbitrary values of $g$, and {\em a two-component 
DS soliton}
\begin{eqnarray}\label{ds}
\phi_1=2\sqrt{\lambda(\lambda-1)} \, \sech(\sqrt{\lambda} z) 
\; , \quad 
u_1=\lambda \, \sech^2(\sqrt{\lambda} z) \; , 
\end{eqnarray} 
which exists only in the integrable case $g=6$. 

To understand what happens for $g \neq 6$, we consider the limit 
$\phi/u\sim \varepsilon \ll 1$ and apply a multi-scale asymptotic analysis. 
In the zeroth order in $\varepsilon$, $\phi=0$ 
and Eq. (\ref{ode}) reduces to a nonlinear equation 
for the component $u(z)$ only, with the supersonic 
\B q soliton solution (\ref{bq}). 
In the first order in $\varepsilon$, we obtain a linear eigenvalue 
problem for $\phi(z)$ characterized by the effective potential $u=u_0(z)$. 
For a given value of $g$, the spectrum of the eigenvalue 
problem consists of $N+1$ discrete eigenvalues 
$\lambda_n = \left(N-n \right)^2$, 
where $n=0, \; 1, \; ... \; N,$ and $N$ is the integer part of 
$(1/2)[\sqrt{1 + (48/g)}-1]$. Each cut-off value 
$\lambda_n$ corresponds to a {\em bifurcation point} of the node-less
scalar soliton $u_0$ where a two-component solution with a  
nonzero component $\phi$ emerges. The latter has $n$ nodes and, near 
the bifurcation point, can be treated as a fundamental (or 
higher-order) bound mode of an effective potential created by the 
soliton $u_0(z)$. The emerging {\em vector soliton} 
 can therefore be characterised by  
 a ``state vector'' $|0,n\rangle$, according to the number of nodes in the 
corresponding components. 

It is easy to see that for $g > 6$ only bifurcations of the 
$|0,0\rangle$ state, which corresponds to the DS soliton 
(\ref{ds}), are possible. First bifurcation of the $|0,1\rangle$
state occurs for the completely integrable 
case $g=6$ at $\lambda_1 = 0$. In this case, the bifurcation pattern 
is identical to that of the completely integrable 
Manakov limit of two coupled NLS equations \cite{manakov}, namely, the 
$|0,0\rangle$ state appears at $\lambda_0=1$, and the 
$|0,1\rangle$ state appears at $\lambda_1=0$. 

{\em Weaker anharmonicity} (smaller $g$) means {\em larger} number of 
possible bound states supported by the effective potential $u(z)$, and 
thus the increasing number of bifurcations. Indeed, the depth of the 
effective trapping potential is inversely proportional to $g$. The $|0,0\rangle$ state always exists, even 
for a shallow potential $u(z)$.

We now consider in detail the formation of {\em multihump solitons} in the 
cases of weak ({\em subcritical}, $g<6$) and strong 
({\em supercritical}, $g>6$) anharmonicity, respectively.

{\em Supercritical regime}. 
In the absence of bifurcating higher-order solutions, the multihump 
solitons are formed only via binding of the $|0,0\rangle$ 
vector solitons. The physics of this mechanism is simple. 
The interaction forces between closely separated fundamental 
solitons are different for both the $\phi$ and the $u$ components. 
Namely, while 
two {\em in-phase} $u$-solitons {\em attract}, the two {\em in-phase} 
$\phi$-solitons {\em repel}. This allows for the existence of multihump 
nodeless modes of the field $\phi$ trapped in the multi-well potential
$u$. Each of such multihump solitons can be considered as {\em a bound 
state of several $|0,0\rangle$ DS solitons}, 
with in-phase humps in both components. 

It is convenient to represent the solution families as 
branches on the {\em bifurcation diagram} $Q$ vs. $\lambda$, where 
$Q\equiv Q_u+Q_\phi\equiv \int u^2 dz +\int \phi^2 dz$ is the total 
soliton power. Typical bifurcation diagram for a supercritical
case ($g=7$) is shown in Fig. \ref{sup_bif}. Solid line represents 
the bifurcating solution $|0,0\rangle$, and it can be seen on the 
close-up of the bifurcation region, that the branches
representing two- and three-hump solutions start off at the same point 
$\lambda=\lambda_0$ but with the energies approximately equal to that 
of two or three $u$-solitons. Examples of such multihump solitons are 
shown in Fig. \ref{sup_prof}, and it is clear that {\em this novel type of
soliton bifurcations occurs from a countable set of infinitely
separated single solitons}. With increasing $\lambda$,
separation between the humps decreases until all solitons of this 
type become single-humped (Fig. \ref{sup_prof}, right column).

\begin{figure}
\setlength{\epsfxsize}{5.2 cm}
\centerline{\epsfbox{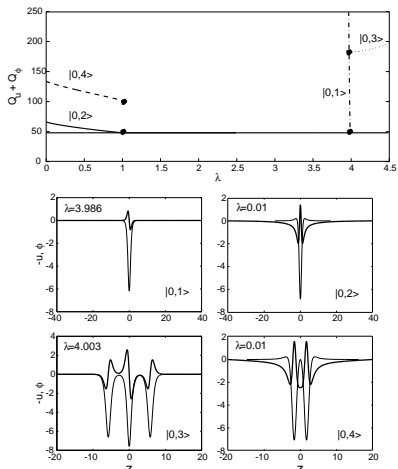}}
\caption{Bifurcation diagram of the system (\ref{ode}) for $g=1$. 
Shown are branches
of the fundamental $u-$soliton (horizontal), $|0,1\rangle$
(dot-dashed), $|0,3\rangle$ (dotted), $|0,2\rangle$ (solid), and
$|0,4\rangle$ (dashed) solitons. Below: profiles of 
$|0,1\rangle$ and $|0,2\rangle$ solitons (top row) and some 
examples of their bound states, 
$|0,3\rangle$=$|0,1\rangle$+$|0,1\rangle$+$|0,1\rangle$ and 
$|0,4\rangle$=$|0,2\rangle$+$|0,2\rangle$ (bottom row).}
\label{sub_bif}
\end{figure}

{\em Subcritical regime}. 
In this case, bifurcations of the $u$-state   
{\em do not lead to multihump solitons}. That is, in 
 sharp contrast to the coupled NLS equations describing vector 
solitons in nonlinear optics, none of the higher-order states 
$|0,n\rangle$ become multihumped in the system under consideration. 
Although the function $\phi$ does have multiple maxima in its intensity
profile, because of the non-self-consistent source for the
$u$-component, it does not cause significant distortions in the shape of the 
effective potential, $u(z)$, and the total intensity
$I(z)=u^2+\phi^2$ remains single-humped. Typical bifurcation diagram
for the case $g=1$ is presented in Fig. \ref{sub_bif}. In this case 
$N=2$, and only bifurcations of the $|0,1\rangle$ (dot-dashed 
line) and $|0,2\rangle$ (solid line) solitons are shown. Corresponding modal 
profiles of the bifurcating solitons are presented in Fig. 
\ref{sub_bif} (top row).
 
Similar to the supercritical regime, the multihump solitons 
can exist only as {\em bound states} of the bifurcating 
$|0,1\rangle$ or $|0,2\rangle$  solitons. They appear at the 
bifurcation points $\lambda_n$, and they have energies equal to a number of
lower-order solitons ``glued'' together by the low-amplitude
components. The number of nodes that the $\phi$-component has in 
the composite soliton depends on the number of $|0,n\rangle$ solitons 
forming that bound state. Typical examples of such solutions are 
presented in Fig. \ref{sub_bif} (bottom row).  

From this analysis, we can conclude that, in this model, the 
multihump solitons appear as bound states of 
$|0,0\rangle$ solitons for {\em any value of anharmonicity parameter}
$g$. In addition, multihump solitons of more sophisticated modal 
structure are also possible.

\begin{figure}
\setlength{\epsfxsize}{6.5 cm}
\centerline{\epsfbox{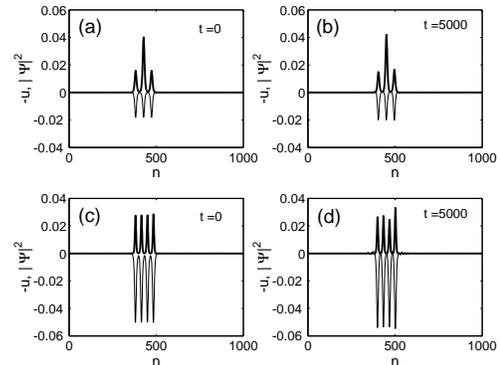}}
\caption{(a,b) Stable dynamics of a three-hump soliton for $g=7$ and  
 $\lambda=0.824$. The family of these solitons is shown in Fig. 
\ref{sup_bif} (dotted line), 
and the corresponding profiles are presented in Fig. 
\ref{sup_prof} (top row). (c,d)  Stable dynamics of a four-hump soliton for 
$g=3.005$ and $\lambda=2.9$.}
\label{sup_dyn}
\end{figure} 

{\em Dynamical stability}.
The second equation of the system (\ref{pde}) is the so-called 
``ill-posed'' (or ``bad'') \B q equation \cite{bad}. 
It possesses an intrinsic 
linear instability and therefore its reliable numerical 
solution for non-zero $\mu$ is unfeasible. This linear instability
is not inherent in the original physical model, and it can be traced to 
neglecting higher-order spatial derivatives in Eq. 
(\ref{pde}). In the case of the energy transport in anharmonic 
molecular chains, Eq. (\ref{pde}) with $\mu=1/12$ originates from 
the following system of discrete equations \cite{Gaididei,Zolotaryuk}
\begin{eqnarray}\label{discr}
i \frac{d}{dt} \Psi_n &=& -\Delta_2(\Psi_n) 
-(W_{n}+W_{n-1}) \Psi_n \; ,\\ \nonumber 
m \frac{d^2}{d t^2} W_n &=& \Delta_2(W_n) + \alpha \Delta_2(W^2_n) 
\\ \nonumber 
&+& \frac{1}{2} (|\Psi_{n+1}|^2 + |\Psi_{n}|^2 - |\Psi_{n-1}|^2 
- |\Psi_{n+2}|^2) \; , 
\end{eqnarray}
where $\Delta_2(X_n) \equiv X_{n+1} + X_{n-1} - 2 X_n $. Discrete 
functions $\Psi_n$ and $W_n$ define, in the continuous limit, the excitation 
wave function, $\psi(x,t)$, and the strain function of the lattice, $w(x,t)$. 

Therefore, it would be justified to study 
the dynamics of the stationary solutions of Eqs. (\ref{ode}) numerically by employing 
the original {\em discrete} dynamical system (\ref{discr}). 
Besides, the argument can be reversed, and such a discrete system 
can be treated as a {\em regularised numerical discretisation scheme} for a 
general system of coupled NLS and ill-posed 
{\B}q equations. 

We investigate the dynamical stability of the multihump solitons 
for two distinct cases of a subcritical and 
supercritical anharmonicity discussed above. The condition of a unit norm 
for the envelope function $\Psi_n$ ($\sum_n |\Psi_n|^2 = 1$) 
is satisfied in all cases,  and $v$ is chosen close to the sound
velocity ($c=0.0244$) to allow for 
a smooth discretisation.

In the {\em supercritical regime} ($g>6$), where multihump solitary waves  
can be formed through binding of several $|0,0\rangle$ states together, 
our numerical simulations indicate that {\em such solitons are stable as 
long as the separation between the humps is sufficiently large}. 
This property is just opposite to that observed for multihump optical
solitary waves \cite{ours}. 
An example of the stable dynamics of a three-hump 
soliton for $g=7$ is shown in Figs. \ref{sup_dyn}(a,b). All solitons
of the DS type, i.e. $|0,0\rangle$ soliton states, 
presented in Fig. \ref{sup_bif} (by solid line) and Fig. \ref{sup_prof} 
(bottom row), exhibit similar stable dynamics.

It is important that the same mechanism of the creation of multihump 
solitons applies to the {\em subcritical regime}. This means that the 
dynamically stable multi-soliton bound states described above exist 
also for $g<6$.  As an 
example, propagation of a four-hump soliton at $g=3.005$ is 
demonstrated in Figs. \ref{sup_dyn}(c,d). After initial adjusting 
of the soliton amplitudes (due to the discretisation), only small 
amplitude breathing  
occurs [see Fig. \ref{sup_dyn}(d)], otherwise the soliton dynamics 
is stable. In contrast, all {\em bifurcating 
higher-order solitons} are dynamically unstable.

Our results on the robustness and stability of multi-soliton states
call for a systematic revision of our understanding of the role of
nonlinear localized modes in a number of physical phenomena related to
the {\em nonlinear transport} in macromolecules \cite{Dav-book} and
even artificial nanoscale structures \cite{prl}, where the coupling of
two (or more) degrees of freedom occurs. {\em How the soliton binding
and existence of multi-soliton states modifies the nonlinear kinetics, nonequilibrium thermodynamics \cite{Leonor}, and other properties of the system}? These questions remain 
to be answered.

In conclusion, we have found robust two-component solitary waves in a
polaron-type model of the energy transport in anharmonic
lattices. We have revealed a novel physical mechanism for the formation of
multihump solitons in a discrete anharmonic lattice and demonstrated their dynamical stability. Along with the recent studies on multihump
optical solitons \cite{moti,ours}, these results call for re-examination of the role of multi-component solitary waves in other 
fields of nonlinear physics.

We thank A.~V.~Zolotaryuk and L. Cruzeiro-Hansson for helpful discussions. 
 S.~M. acknowledges support from the Australian Research Council. The work of E.~O. and Yu.~K. is supported by the Performance and Planning Fund of The Australian National University.

\end{multicols}
\end{document}